\documentclass[aps,prl,showpacs]{revtex4}

\newcommand{\be}{\begin{equation}}
\newcommand{\ee}{\end{equation}}

\begin{document}

\large

\title{Group-Theoretical Derivation of Aharonov-Bohm Phase Shifts}

%\draft
%voffset=.5in
%@twocolumnfalse\endcsname
\author{C. R. Hagen\footnote{mail to hagen@pas.rochester.edu} }
\affiliation{Department of Physics and Astronomy\\
University of Rochester\\
Rochester, N.Y. 14627-0171}

\begin{abstract}
The phase shifts of the Aharonov-Bohm effect are generally determined by means of the partial wave decomposition of the underlying Schr\"{o}dinger equation.  It is shown here that they readily emerge from an $\mathfrak{o} (2,1)$ calculation of the energy levels employing an added harmonic oscillator potential which discretizes the spectrum.
\end{abstract}

%\begin{keyword}

%\end{keyword}
%\end{frontmatter}

\maketitle

%\noindent DRAFT TeXed on \today
%\vskip 1cm

% main text
The Aharonov-Bohm (hereafter simply AB) effect \cite{effect}is one of the most intriguing predictions of quantum mechanics inasmuch as it implies a non-vanishing cross section for a scattering process  in which there is no classical force  acting on the particle.  The relevant system is described  by the interaction of a charged particle of mass $M$ and energy $E$ with a magnetized filament whose  vector potential is of the form
$$eA_i=\alpha {\overline x}_i/r^2$$
where $r^2\equiv x_i x_i$ and ${\overline x_i}\equiv \epsilon_{ij}x_j$ with $\epsilon_{ij} $ the usual Levi-Civita tensor  in the two dimensional space of the Cartesian coordinates $x_i$.    The Schr\"{o}dinger equation for this system is readily brought to the form
$$  \left[ {1\over r} {d \over {dr} }  r {d \over {dr} }  + k^2- { (m+\alpha)^2\over {r^2} } \right] f_m(r)=0$$
where $k^2=2ME$ ($\hbar=1$) and the wave function $\psi (r,\phi)$
has been expanded as
$$  \psi(r,\phi) = \sum_{-\infty}^{\infty} e^{im\phi}f_m(r)  . $$
In the standard way one can extract from the Bessel function solutions of the radial equation an expression for the phase shifts associated with the partial wave $m$.  This leads to the well known result 
$$\delta_m = - {\pi \over 2}[|m+\alpha|-|m|].$$
It is the object of the present work to show that this result for the phase shifts can be obtained also from strictly group theoretical methods with no reference to the underlying wave equation.    

The most straightforward way to achieve this goal is to discretize the spectrum of the system.  That is to say, one circumvents the complications associated with the fact that all positive energy states can be realized for the AB problem by the inclusion of an attractive harmonic oscillator potential so that one is dealing with a discrete rather than a continuous spectrum.   In the limit in which this additional potential goes to zero one regains the original system  while extracting the phase shifts as the phase of the corresponding Fredholm determinant.

The Hamiltonian of the modified system is given by 
$$H = {1\over 2M}  ( {\bf p} - \alpha {\overline {\bf x} }/ r^2 )^2 + {1\over 2} M {\omega}^2 r^2 $$
where $p_i$ and $x_i$ satisfy the commutation relation
$$[x_i,p_j] = i \delta_{ij}.$$
Upon rescaling $x_i=Q_i / \sqrt {M \omega}$, $p_i=P_i \sqrt {M  \omega}$ 
and defining $J_3\equiv H/2\omega $ there follows
$$J_3={1\over 4}(P_i-\alpha \overline{Q}_i/Q^2)\delta_{ij} (P_j-\alpha\overline{Q}_j/Q^2)+{1\over 4}Q^2.$$  
 Making use of the decomposition of the identity according to 
$$\delta_{ij}=  (  Q_iQ_j + {\overline Q}_i {\overline Q}_j)/ Q^2,$$
$J_3$ is brought to the form
$$J_3={ 1\over 4}
 \left[ {\bf P} \cdot {\bf Q} {1\over{Q^2} }  {\bf Q} \cdot {\bf P} + {  (L+\alpha)^2 \over {Q^2} } + Q^2 \right]$$
where the angular momentum operator $L\equiv Q_i \cdot {\overline{P}}_i$ has been introduced.  The definition of two additional operators
$$K_1={1\over4}[{\bf P}\cdot {\bf Q}{1\over{Q^2}}{\bf Q}\cdot{\bf P}+{(L+\alpha)^2\over{Q^2}}-Q^2]$$
and 
$$K_2={1\over4} ( {\bf P}\cdot {\bf Q}+{\bf Q}\cdot{ \bf P} ) $$
leads to a closed set of commutators for the operator set $J_3$, $K_1$, and $K_2$.  Specifically, one finds that 
$$[J_3,K_1] = iK_2$$
$$[J_3,K_2] = - iK_1$$
and 
$$[K_1,K_2 ]= - iJ_3.$$
These three operators thus comprise an $o(2,1)$ algebra whose Casimir operator
 $C \equiv J_3^2 - K_1^2-K_2^2$ is given by 
\be
C = { (L+\alpha)^2 - 1\over 4}.
\ee
It should be noted that this algebra was also considered for the $\alpha=0$  case as a method to obtain the spectrum of the isotropic harmonic oscillator by Bacry and Richard\cite{Bacry}.

The spectrum of $L$ is, of course, that of the two dimensional rotation group.  Since the representations are one-dimensional, they are characterized by the single number $m$ which is necessarily integral for single valued representations.  Thus the problem of obtaining the spectrum of the modified AB system is essentially to determine the relevant representations of $\mathfrak {o}(2,1)$.  These have been tabulated by Barut and Fronsdal \cite{BF}, extending an earlier work of Bargmann \cite{Bargmann}.  They showed  that there are four series of representations of 
$\mathfrak {o}(2,1)$ characterized by a pair of complex numbers $(\Phi, E_0)$. The series are concisely summarized in a recent work by Friedmann and the author \cite{author}.  The analysis of that work is readily adapted to the problem under consideration here with the conclusion again being that only the representation $D^+(\Phi)$ with $E_0=-\Phi$ and $J_3 $ spectrum $E_0+n$ where $n=0,1,2,...$ is an admissible candidate.  It furthermore must be restricted by the conditions $Im(E_0)=0$ and $\Phi<0$ in order to satisfy both the requirement of unitarity as well as  the condition on the positivity of the spectrum of $J_3$.  The Casimir operator in this case is given by 
$$C=E_0(E_0-1)$$
which upon comparison with (1) yields that 
$$E_0={1\over2}[1\pm |m+\alpha|].$$
Thus the spectrum of $H/\omega \equiv 2J_3$ is given by 

$$ {1\over \omega}E_n=2n+1\pm | m+ \alpha |.$$
As shown in the Appendix only the upper sign is compatible with the positivity of the Hamiltonian and thus the above reduces to 
$${1\over \omega}E_n=2n+1+| m+\alpha |.$$

This derivation of the spectrum of the Hamiltonian now allows one to infer the  AB phase shifts.  To this end one employs the relation between the Fredholm determinant $D_m(E)$ and the corresponding phase shift $\delta_m$.  It has been shown by DeWitt \cite {DeWitt} that this is given by 
$$\exp[2i \delta_m(E)]=D_m(E+i0)^*/D_m(E+i0)$$
in the limit in which the confining oscillator potential is reduced to zero.  Using the identity
$$\log \det O=Tr \log O,$$
there follows
$$\log D_m=Tr [\log(E-H)-\log(E-H_{\alpha=0} ) ]$$
whence 
%\begin{eqnarray*}
$$
\log D_m  =    \sum_{n=0}^{\infty}\left [ \log [E-\omega(2n+1+ | m+\alpha|) ]
 -  \log [ E-\omega(2n+1+|m|) ] \right] .
%\end{eqnarray*}
$$
In the limit of small $\omega$ this becomes 
%\begin{eqnarray*}
$$
\log D_m  =  { 1\over2\omega} \int_{0}^{\infty} dx\left[ \log(E-x-\omega | m+\alpha | ) 
 -  \log(E-x-\omega | m | ) \right],
$$
%\end{eqnarray*}
the imaginary  part of which yields
$$e^{2i\delta_m}=\exp\{-i\pi [ | m + \alpha | - | m | ]\}$$
and the immediate identification of the AB phase shifts as
$$\delta_m=-{\pi \over 2} [ | m+ \alpha | - | m | ].$$
Thus the AB phase shifts have been obtained solely by group theoretical methods making no reference to the usual wave equation approach.

\section{Appendix}

In this appendix it is shown that only the positive sign for the $|m+\alpha|$ term in the spectrum of $J_3$ is allowed.  To this end one defines
$$\langle J_3\rangle_{m_0}\equiv{\langle m_0|J_3|m_0\rangle\over \langle m_0|m_0 \rangle}.$$
Upon variation of $J_3$ there follows
%\begin{eqnarray*}
$$
\delta \langle J_3\rangle_{m_0}   = 
{ \langle \delta m_0|J_3|m_0 \rangle + \langle m_0|J_3|\delta m_0 \rangle \over \langle m_0|m_0\rangle} 
 - { \langle \delta m_0|m_0\rangle + \langle m_0|\delta m_0\rangle \over \langle m_0|m_0\rangle } 
 \langle J_3\rangle_{m_0}  + {\langle m_0|\delta J_3|m_0 \rangle \over\langle m_0|m_0\rangle}.
%\end{eqnarray*}
$$
Using the fact that $| m_0 \rangle$ is an eigenvector of $J_3$ with eigenvalue $m_0$ this reduces to 
$$\delta \langle J_3 \rangle_{m_0} = {\langle m_0|\delta J_3|m_0 \rangle \over \langle m_0 |m_0 \rangle}.$$
Thus a variation of $J_3$ associated with the parameter $|m+\alpha |$ yields 
$${\delta m_0\over \delta (m+\alpha)^2} =
{1\over 4} {\langle m_0|Q^{-2}|m_0 \rangle\over \langle m_0|m_0\rangle}\geq 0.$$
Comparison with the double-valued result for $m_0$, however, yields
$${\delta m_0\over (m+\alpha)^2}=\pm {1\over 4}{1\over |m+\alpha|}$$
thereby establishing the inadmissibility of the lower sign solution.  It is important to note in obtaining this result that the matrix element $\langle Q^{-2}\rangle$ must be finite and positive inasmuch as all three terms which comprise the operator $J_3$ are themselves  positive definite.

\end{document}